\documentclass[12pt,fleqn]{article}
\usepackage{verbatim,amssymb,amsmath,graphics,amsthm,cite}
\numberwithin{equation}{section}

\newtheorem{definition}{Definition}[section]
\newcommand{\bs}{\boldsymbol}

\newcommand{\M}{{\cal M}}
\def\P{{\cal P}}
\newcommand{\T}{{\cal T}}

\begin{document}
\title{A Note on the Topology of Space-time in Special Relativity} 
\author{
S.~Wickramasekara\footnote{sujeewa@physics.utexas.edu}\\ 
Department of Physics, University of Texas at Austin\\
Austin, Texas 78712}
\date{}
\maketitle
\begin{abstract}
We show that a topology can be defined in the four dimensional
space-time of special relativity so as to obtain a topological
semigroup for time. The Minkowski 4-vector character of space-time
elements as well as the key properties of special relativity 
are still the same as in the standard theory. However, the new
topological structure allows the possibility of 
an intrinsic asymmetry in the time evolution of physical systems.\\ 
\end{abstract}
\section{Introduction}\label{I} 
It is the received point of view that Einstein's special theory of
relativity is above all a theory of time. In some sense, the theory
``unifies'' space and time into a single entity, a four dimensional
space-time, and a physical event is cataloged by a
point $(x_0,\bs{x})$ in this space-time. The new notion of time,
though quite different from the preceding Newtonian time, still has the
reversible, symmetric property of the latter. More precisely, time in
relativity is assumed to be modeled by the Euclidean real line, and as
such is reversible and symmetric in the sense that the Euclidean real
line is a Lie group under addition. Within its
topological and algebraic structure, there is no
natural way to define a flow of time in the four dimensional
space-time of special relativity. 

On the other hand, the time evolution of most macroscopic systems 
has an amply manifest irreversible character, often associated 
with the second law of thermodynamics. Moreover, it is believed
that the time 
evolution of certain microphysical processes, such as resonance
scattering and the decay of elementary particles, also possesses such
an irreversible nature. Various attempts have
been made to develop a quantum theory which accommodates the time
asymmetry of such phenomena~\cite{bohm1,gadella1,rgv}, and even the
asymmetry in the time evolution of the universe as a
whole~\cite{gell_mann, castagnino}. Perhaps the most significant
feature common to all these developments is that the asymmetry in time
evolution is attempted to be obtained as a property of its
representation in the space of sates of the system. This is
particularly notable in~\cite{bohm1,gadella1,rgv} where the
asymmetric time evolution of the microphysical system is realized by
way of a {\em semigroup} of continuous linear operators defined in
a suitably constructed rigged Hilbert space.

What is implicitly assumed in these theories is that the time which
gets {\em represented asymmetrically} is still the time of special
relativity, and as such, has no manifest irreversibility or asymmetry
at the space-time level. That is, the evolution parameter associated
with the macroscopic apparatuses which characterize the states and
observables appertaining to a quantum physical system is still taken
to be the classical time of special relativity, reversible and
static; it is the evolution parameter of the states and observables of
the quantum physical system that acquires an
asymmetry. 

The main technical result we present in this paper is   
a topological structure for the space-time of special
relativity that allows for asymmetric time evolutions. Further,  
this topological structure is introduced in 
a manner completely consistent with the tenets of
special relativity -- in particular, without contradicting any of the
experimental tests confirming special relativity. However, the new topology
provides a better framework for the time-asymmetric 
quantum theories, such as those developed
in~\cite{bohm1,gadella1,gell_mann,castagnino,rgv}, in that it    
endows the structure of a topological semigroup
on the set of space-time translations and 
consequently leads to a {\em Poincar\'e semigroup} for relativistic
symmetries (and asymmetries). 

\section{A Topology for Space-time}\label{II}
The space-time of special relativity is assumed to be a four
dimensional manifold, $M$. It has the topology of ${\mathbb{R}}^4$,
the four dimensional Euclidean space. A metric tensor $g_{\mu\nu}$ is
also defined on $M$, for the ``length'' preserved under Lorentz
transformations is not that which is compatible with the Euclidean
topology. In fact, what is at the heart of all experimentally
observable predictions of special relativity is this metric structure,
and not the topological structure of $M$, which is 
Euclidean. 

It is perhaps the case that the mathematical structure of a physical
theory, especially its topology, is never completely
determined by the physics of the processes it seeks to
describe. A topological structure involves such notions as local bases
at a point and infinite sequences which are indeterminable by physics
alone because the totality of experimental information is never
``complete'': The number of experiments that can be performed is
finite; the experimental apparatuses allow the measurement of only
finite quantities--not infinitesimals; and no measurement is without
error. The theoreticians use the freedoms resulting from this
incompleteness to construct the mathematical structure of a theory in
a way that is not necessarily dictated completely by physics, and how
these freedoms are utilized in a given theory is to be valued on the
grounds of its overall success. Indeed, the absence of the uniqueness
of the topology of a physical theory is an interesting query in its
own right.  A case in point is the Euclidean topology of $M$: since it  
involves open sets (or infinite sequences), just as any other
topological structure, this Euclidean topology 
cannot be deduced from experiments alone. 
In addition, it is not the most convenient one to use when describing
certain phenomena, such as time asymmetric processes.  
Here, we propose to alter the topological structure of $M$
while leaving intact the (algebraic) properties of how an element of
$M$ transforms under Lorentz transformations.

Before we introduce this topology, it is worthwhile to recall a few well
known definitions and notions from the theory of groups:

\noindent{\bf Groups} A group is a set $G$ with an operation
$G\otimes G\rightarrow G$, denoted by $(a,b)\rightarrow ab$, 
such that
\begin{itemize}
\item[a)] The operation is associative, i.e., $(ab)c=a(bc)$.
\item[b)] There is an identity element $e$ in $G$ such that
$ea=ae=a$ for all $a\in G$.   
\item[c)]
For every $a\in G$, there exits its inverse $a^{-1}$ such that
$aa^{-1}=a^{-1}a=e$. 
\end{itemize}
Notice that this operation, often called multiplication, 
imposes an {\em algebraic} structure on the set $G$; it is not a
topological structure.

\noindent{\bf Topological Groups} A topological group is a group in which a
topology is defined so as to make the above group operations
continuous. That is, for each $a\in G$, the mappings $x\rightarrow ax$
and $x\rightarrow xa$ are homeomorphisms of $G$ onto $G$. So is the 
mapping $x\rightarrow x^{-1}$. These continuity requirements can be
more concisely stated by way of the continuity of the mapping $f:\
G\otimes G\rightarrow G$ defined by
\begin{equation}
f(a,b)=ab^{-1}\label{0.1}
\end{equation}
It is clear that such a topology on $G$ is completely determined by
any local base at the identity element $e$ of $G$.

\noindent{\bf Lie Groups} A topological group $G$ is called a Lie group if its
topology is such that $G$ is a differentiable manifold and the mapping
\eqref{0.1} is $C^\infty$. Many symmetry transformations in physics
are assumed to constitute Lie groups; the Poincar\'e group and
symmetry groups of particle physics 
such as $SU(3)$ are examples.
privilege
Along with these well known classical concepts, we are in need of
the notion of topological semigroup. For the purposes of this paper,
we define it as follows:\\
\noindent{\bf Topological Semigroups}
A topological semigroup is a topological space $S$ with an internal
operation $S\otimes S\rightarrow S$, denoted by $(a,b)\rightarrow ab$,
such that
\begin{itemize}
\item[a)] The operation is continuous.
\item[b)] It is associative, i.e., $(ab)c=a(bc)$.
\item[c)] There is an identity element $e$ in $S$ such that
$ea=ae=a$ for all $a\in S$.   
\end{itemize}
Thus every topological group is a topological semigroup; of course,
what is of interest here is  those semigroups that are not topological
groups. 
 
After these introductory remarks, we now introduce the central
idea of this paper.

\begin{definition}
Consider the collection of intervals of the form $[a,\ b),\
a,b,\in{\mathbb{R}}$, where ${\mathbb{R}}$ is the real line. It is
clear that these sets provide a base for a  topology for
${\mathbb{R}}$. Let $\tilde{\mathbb{R}}$ denote the set of real
numbers endowed with this topology.
\end{definition}

It is easy to verify that the mapping
$\tilde{\mathbb{R}}\otimes\tilde{\mathbb{R}}
\rightarrow\tilde{\mathbb{R}}$ 
defined by $(t_1,t_2)\rightarrow t_1+t_2$ is continuous. However, the mapping
$t\rightarrow -t$ is not continuous on $\tilde{\mathbb{R}}$. This
means $\tilde{\mathbb{R}}$ is a topological semigroup --not a Lie
group-- under the operation of addition. Further, in contrast to
$\mathbb{R}$, $\tilde{\mathbb{R}}$ is not locally compact.  

We want to propose $\tilde{\mathbb{R}}$ as the mathematical image of
time. It is interesting to notice that $\tilde{\mathbb{R}}$ is an
algebraic group, and thus a notion of past can still be defined by way
of the mapping $t\rightarrow-t$. Since this mapping is not continuous,
however, $\tilde{\mathbb{R}}$ does not have the reversible (i.e., Lie
group) character that time
acquires when modeled by the usual Euclidean line $\mathbb{R}$. Now we
may define our space-time:

\begin{definition}
Consider the direct product space
$\tilde{\mathbb{R}}\otimes{\mathbb{R}}^3$, where 
${\mathbb{R}}^3$ is the usual three dimensional Euclidean
space. Define a topology on $\tilde{\mathbb{R}}\otimes{\mathbb{R}}^3$
by declaring sets of the form 
$V_1\otimes V_2$ open when $V_1$ is open in $\tilde{\mathbb{R}}$ and
$V_2$, in  ${\mathbb{R}}^3$. Let $\M$ be the space
$\tilde{\mathbb{R}}\otimes{\mathbb{R}}^3$ endowed with this product
topology, and let $\tau_{\M}$ denote the topology itself.
\end{definition}

\subsection{The Semigroup of Space-time Translations}\label{III}
$\M$ can be made into (an algebraic) vector space of operators acting
on itself. To that end, let $(a_0,\bs{a})$ be an element of
$\M$. Then, for all $x\equiv(x_0,\bs{x})\in\M$, the mapping
\begin{equation}
(a_0,\bs{a}):\ (x_0,\bs{x})\rightarrow(x_0+a_0,\bs{x}+\bs{a})\label{1}
\end{equation}
defines the desired action on $\M$. It is clear that \eqref{1} is
$\tau_{\M}$-continuous on $\M$. The multiplication on the set
of operators $\{(a_0,\bs{a})\}$ defined by means of composition
\begin{equation}
(a_0,\bs{a})(b_0,\bs{b})=(a_0+b_0, \bs{a}+\bs{b})\label{2}
\end{equation}
is continuous with respect to $\tau_\M$. Furthermore, the inverse
mapping 
\begin{equation}
(a_0,\bs{a})\rightarrow(a_0,\bs{a})^{-1}=(-a_0,-\bs{a})\label{2a}
\end{equation}
is not continuous in the topology $\tau_\M$. Therefore, we see that
$\M$ acquires the structure 
of a topological semigroup (of operators on $\M$ itself). We shall
refer to this semigroup as the semigroup of space-time translations,
or simply as the translation semigroup, $\T$. 

\subsection{Lorentz Transformations}\label{IV}

Let $B$ be the open unit ball in $\mathbb{R}^3$, i.e.,
\begin{equation}
B=\{\bs{v}:\; \bs{v}\in{\mathbb{R}}^3,\ |\bs{v}|<1\}\nonumber
\end{equation}
For every $\bs{v}\in B$, we may define a linear operator
$\Lambda(\bs{v})$ on $\M$ by way of the equality

\begin{equation}
\Lambda(\bs{v})
\left(
\begin{array}{c}
x_0\\
\bs{x}
\end{array}
\right)=
\left(
\begin{array}{c}
{\gamma(x_0-\bs{v}.\bs{x})}\\
{\bs{x}+\frac{\gamma-1}{v^2}(\bs{v}.\bs{x})\bs{v}-\gamma\bs{v}x_0}
\end{array}
\right)
\label{2.2.1} 
\end{equation}
where $\gamma=\frac{1}{\sqrt{1-v^2}}$. As a matrix, the
$\Lambda(\bs{v})$ has the  form
\begin{equation}
\Lambda(\bs{v})=
\left(
\begin{array}{cccc}
\gamma&-\gamma v_1&-\gamma v_2&-\gamma v_3\\[0.2cm]
-\gamma
v_1& 1+\frac{\gamma-1}{v^2}v_1^2&\frac{\gamma-1}{v^2}v_1v_2 
&\frac{\gamma-1}{v^2}v_1v_3\\[0.2cm]
-\gamma
v_2& \frac{\gamma-1}{v^2}v_2v_1&1+\frac{\gamma-1}{v^2}v_2^2
&\frac{\gamma-1}{v^2}v_2v_3\\[0.2cm]
-\gamma
v_3& \frac{\gamma-1}{v^2}v_3v_1 &\frac{\gamma-1}{v^2}v_3v_2
&1+\frac{\gamma-1}{v^2}v_3^2
\end{array}
\right)\label{2.2.2}
\end{equation}
where $\bs{v}=(v_1,v_2,v_3)$. Both \eqref{2.2.1} and \eqref{2.2.2} are
well known from the standard theory: $\Lambda(\bs{v})$ is just the
familiar Lorentz boost operator on the space-time manifold $M$. Thus,
algebraically, the boost operators (defined by \eqref{2.2.1} or
\eqref{2.2.2}) on the space-time
$\M$ are identical to the
those on the conventional space-time $M$ of special
relativity. However, the space-time is now endowed with a
different topology $\tau_\M$, and we must verify that the operators
$\Lambda(\bs{v})$ are continuous with respect to $\tau_\M$. 

Recall that an operator $A$ defined on a topological space $S$ is said
to be continuous if for every open set $U$ of $S$ there exists another
$W$ such that $A(W)\subset U$. Further, in the present case it is
sufficient to consider a 
boost operator of the form $\Lambda(v_1)$, for which \eqref{2.2.1}
reduces to
\begin{equation}
\Lambda(v_1)(x_0,\bs{x})=(\gamma(x_0-v_1x_1),\gamma(x_1-v_1x_0),x_2,x_3)
\equiv(x'_0,\bs{x'})\label{2.2.3}
\end{equation}
where $v_1\equiv\bs{v}=(v_1,0,0)$. Next, let $U$ be an open
neighborhood of $(x'_0,\bs{x'})$ of the form $[x'_0,
x'_0+\epsilon)\otimes(x'_1-\epsilon,x'_1+\epsilon)\otimes V$, 
where
$V=(x'_2-\epsilon,x'_2+\epsilon)\otimes(x'_3-\epsilon,x'_3+\epsilon)$. 
It then follows from \eqref{2.2.3}, which shows that $\Lambda(v_1)$
acts on the coordinates $x_2$ and $x_3$ as the identity, that any
neighborhood $W$ of $(x_0,\bs{x})$ of the form $[x_0,
x_0+\delta)\otimes(x_1-\delta,x_1+\delta)\otimes V$, where
$\delta<\epsilon\sqrt{\frac{1-v_1}{1+v_1}}$, fulfills the relation
\begin{equation}
\Lambda(v_1)(W)\subset U\label{3}
\end{equation}
Therefore, $\Lambda(v_1)$ is a continuous operator on $\M$.

It is obvious that the rotation operators $R(\bs{\theta})$ are also
continuous on $\M=\tilde{\mathbb{R}}\otimes{\mathbb{R}}^3$ as they act
non-trivially only on ${\mathbb{R}}^3$.

Now, let $L=\{\Lambda\}$ be the totality of the boost operators
$\Lambda(\bs{v})$ and rotation operators $R(\bs{\theta})$. As in
the standard theory, under the multiplication defined by usual
composition of operators, $L$ is a Lie group--the well known
homogeneous Lorentz group. We have shown that it is a group of
continuous operators on the new space-time $\M$.

\subsection{Poincar\'e Semigroup}\label{V}
Consider the semidirect product of the translation semigroup 
$\T$ with the Lorentz group$L$. Following the common
practice, we denote elements of this semidirect product set $\P$ by
$(\Lambda,a)$, where $\Lambda\in L$ and $a\in\T$. From the
considerations of Sections~\ref{III} and \ref{IV}, we see that
$(\Lambda, a)$ is a continuous operator on $\M$, defined by
$(\Lambda,a)x=\Lambda x+a$. As usual, we define a product rule on
$\P$ by the composition of operators:
\begin{equation}
(\Lambda_1, a_1)(\Lambda_2, a_2)=(\Lambda_1\Lambda_2,
a_1+\Lambda_1a_2)\label{4}
\end{equation}
This is an associative multiplication on $\P$ under which the set
remains closed. Furthermore, for each $(\Lambda, a)\in\P$, there
exists an inverse element $(\Lambda,a)^{-1}$ given by,
\begin{equation}
(\Lambda,a)^{-1}=(\Lambda^{-1}, -\Lambda^{-1}a)\label{5}
\end{equation}
Thus, under the product rule~\eqref{4}, $\P$ acquires the structure of
an {\em algebraic} group.

Consider now the topological properties of $\P$. From \eqref{2} and
\eqref{3}, we see that \eqref{4} is a continuous mapping of
$\P\otimes\P$ into $\P$. However, \eqref{2a} implies that
\eqref{5} is not continuous on $\P$. This means that the multiplication
defined by \eqref{4} turns $\P$ into a {\em topological semigroup}
with respect to the new topology $\tau_{\M}$ we have introduced on
space-time. Recall that under the usual Euclidean topology, the
multiplication~\eqref{4} makes $\P$ a Lie group, the very well known
Poincar\'e group. We still retain
the algebraic structure of the Poincar\'e group, but introduce here a
topology that makes $\P$ only a semigroup. We call this
topological semigroup the Poincar\'e semigroup. 
\section{Concluding Remarks}
That the set of translations and Lorentz transformations on the
relativistic space-time forms a Lie group is a mathematical
assumption--one perhaps intrinsically incapable of being verified by
direct experiments. Among the conclusions to which this assumption
leads is the necessarily reversible, unitary time evolution (in the
Hilbert space representation) of quantum
physical systems. However, as
pointed out in the Introduction, there exist many physical processes
the time evolution of which is not unitary and reversible. Although
attempts 
have been made to construct quantum physical theories to describe
these phenomena, how these theories are to be reconciled with the
structure of relativistic space-time (in particular, the Lie group
structure of space-time translations) has not been studied.

This brief note investigates the possibility of endowing the four
dimensional space-time of special relativity with a topology 
which is different from its usual Euclidean topology. We
still retain the key properties of special relativity, all of which
originate from the algebraic structure of Lorentz transformations. The
new topology, however, allows us to view time in way that is quite
different from the static character it assumes in orthodox
special relativity. As immediate consequences of the new topological
structure, we have shown that space-time translations on $\M$ define a
topological semigroup, whereupon we obtained the Poincar\'e semigroup
for the set of relativistic transformations on $\M$. This structure
may be the one 
that provides a proper context and framework for the time asymmetric
quantum theories, such as those developed in some of the works cited below.
It now remains to investigate the implications of this topology on the
fundamental equations of physics  
--i.e., the meaning of partial derivatives with respect to time needs to
be explored-- and to construct the representations of the Poincar\'e
semigroup in the space of states of quantum mechanical systems. We
shall undertake these tasks in a forthcoming paper.

\end{document}